# Space-Division Multiplexing for Microwave Photonics

***Sergi García, Rubén Guillem, Mario Ureña and Ivana Gasulla***

ITEAM Research Institute, Universitat Politècnica de València, Valencia, Spain

Author e-mail address: ivgames@iteam.upv.es

***Abstract:*** *Despite space-division multiplexing fibers were initially envisioned as pure transmission media, they exhibit as well great potential for signal processing. We review here different solutions we developed for Microwave signal processing devices and links.*

## 1. INTRODUCTION

The application of Space-Division Multiplexing (SDM) technologies in optical fibers has been welcomed as a promising approach to overcome the upcoming capacity crunch of conventional single-mode fiber (SMF) communications, [1]. Despite SDM fibers were initially envisioned as parallel distribution media for core and metro optical networks, they can actually be applied to a wide range of scenarios where, in addition, they exhibit the potential not only for signal distribution but also for signal processing. In this regard, recent research activities have focused on parallel distribution in radio-over-fiber access networks and multiple antenna connectivity [2], radiofrequency (RF) signal processing [3] as well as multi-parameter fiber sensing. In particular, we foresee that microwave photonics (MWP) signal processing [4] can benefit greatly from the use of different SDM fiber media in terms of compactness and weight, while assuring broadband versatility, reconfigurability, and performance stability.

For RF signal processing functionalities such as microwave signal filtering, antenna beam-steering for phased-array antennas and arbitrary waveform generation [4], the propagated signals must exhibit different group delay and/or chromatic dispersion characteristics. Nowadays, the trend is to dedicate a given system (either a fiber-based system or a photonic integrated circuit) to process the signal and a separate optical fiber link to distribute it. There is however one revolutionary approach that has been left untapped in finding new ways to address this challenge: exploiting *space*, the last available degree of freedom for optical multiplexing. We have recently proposed to implement MWP parallel signal processing functionalities while we distribute the data signal to the end user. This challenge implies the development of innovative multicore fiber (MCF) or few-mode fiber (FMF) solutions where the different spatial paths (cores or modes) translate into the different propagation characteristics (in terms of group delay and chromatic dispersion) required for signal processing. The concept of "fiber-distributed signal processing" shows great potential, among others, in the field of future converged fiber-wireless telecommunications networks, as those envisioned for 5G communications and the Internet of Things. Figure 1(a) shows, as an example, a typical fiber-wireless communications scenario where the use of SDM technologies brings an important reduction in terms of space, weight and power consumption as compared to using a bundle of standard single-mode fibers.

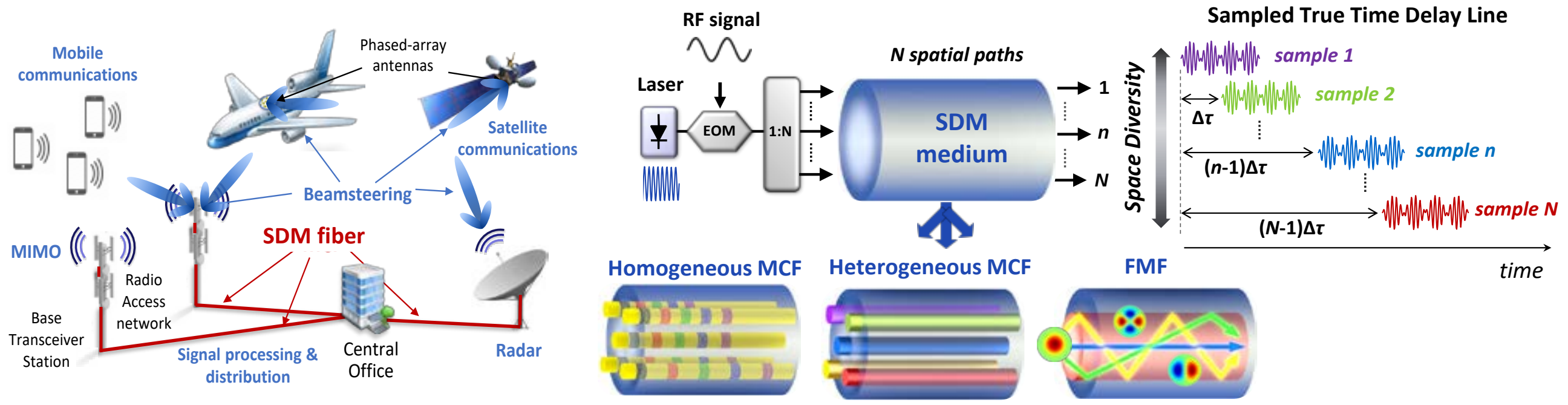

Fig. 1 (a) Typical Microwave photonics fiber-wireless communication scenarios; (b) Generic sampled delay line based built upon SDM fibers.

This research was supported by the ERC Consolidator Grant 724663, the Spanish MINECO Project TEC2016-80150-R, the Spanish scholarship MINECO BES-2015-073359 for S. García and the Spanish MINECO Ramon y Cajal fellowship RYC-2014-16247 for I. Gasulla.

## 2. SAMPLED DELAY LINE OPERATION BASED ON SDM FIBERS

At the heart of most MWP applications, we can find the true time delay line (TTDL), an optical subsystem that provides a frequency independent and tunable delay within a given frequency range, [4]. Let's see how to implement this processing building block with a generic SDM medium. As Fig. 1 (b) depicts, the goal is to obtain at the output of our link (or component) a set of time-delayed replicas of the modulated signal. This series of samples must feature a constant differential delay between them (named as basic differential delay, $\Delta\tau$), [3]. If only one optical wavelength is implicated, as shown in Fig. 1 (b), the TTDL features 1D performance where all the samples result from exploiting the fiber spatial diversity. But this approach offers in addition 2D operation if we combine the spatial diversity with the optical wavelength diversity provided by the use of multiple optical wavelengths. We can implement a sampled TTDL using a heterogeneous MCF if each core features different group delay and dispersion characteristics [5], a homogeneous MCF with certain dispersive elements inscribed along the cores [6] or even few-mode fibers [7]. We summarize next the advances accomplished in each one of these technologies.

### *A. Heterogeneous Multicore Fibers*

While digital signal distribution usually demands similar propagation characteristics in all the cores, the implementation of tunable TTDLs requires different group delay values for a specific wavelength. This requirement calls for the development of customized heterogeneous MCFs where, while keeping a low intercore crosstalk, we can tailor the chromatic dispersion of each single core. The group delay $\tau_n(\lambda)$ of core $n$ can be expanded following a 2$^{nd}$-order Taylor series around a reference or anchor wavelength $\lambda_0$ as: $\tau_n(\lambda) = \tau_n(\lambda_0) + D(\lambda - \lambda_0)$, where $D_n$ is the chromatic dispersion of core $n$ at the anchor wavelength. For proper tunable time delay operability, we need a linear increment of $D_n$ with the core number $n$. If all cores share the same $\tau_n(\lambda_0)$, we can control the basic differential delay, $\Delta\tau = \tau_{n+1}(\lambda) - \tau_n(\lambda)$, to achieve continuous tunability from 0 up to tens (or even hundreds) of ps/km. This allows for the implementation of distributed signal processing on the fly for link lengths up to a few kilometers.

The design of the required heterogenous MCF involves the customization of the refractive index profile of each single core, where trench-assisted configurations are preferred since they provide more design versatility. Different designs of heterogeneous 7-core fibers behaving as sampled TTDLs have been reported. In [5], we presented an optimum design in terms of both group delay tunability and minimum crosstalk for a fiber whose schematic cross section is depicted in Fig. 2(a). The computed parameters for each core are gathered in Fig. 2(b), where $a_1$ is the core radius, $a_2$ is the core-to-trench distance, $w$ is the trench width and $\Delta_1$ is the core-to-cladding relative index difference. This MCF offers a linear spectral group delay with incremental values from core to core, as shown in Fig. 2(c). In other words, it fulfils the demands for tunable delay line operation and can be used as the basic element to perform different signal functionalities, as signal filtering or radio beamsteering for phased array antennas.

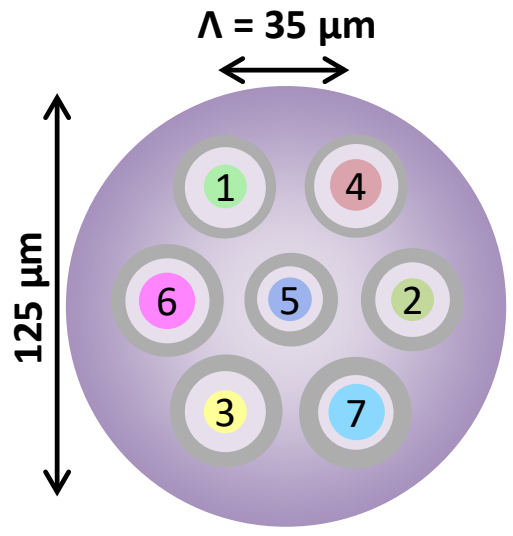

| Core | $a_1$ (µm) | $\Delta_1$ (%) | $a_2$ (µm) | $w$ (µm) | $D$ (ps/km/nm) |
|---|---|---|---|---|---|
| 1 | 3.42 | 0.3864 | 5.48 | 3.02 | 14.75 |
| 2 | 3.60 | 0.3762 | 5.03 | 2.61 | 15.75 |
| 3 | 3.62 | 0.3690 | 4.35 | 3.32 | 16.75 |
| 4 | 4.26 | 0.3588 | 4.92 | 4.67 | 17.75 |
| 5 | 3.49 | 0.3476 | 2.81 | 5.41 | 18.75 |
| 6 | 4.79 | 0.3435 | 3.35 | 3.32 | 19.75 |
| 7 | 4.98 | 0.3333 | 2.42 | 4.05 | 20.75 |

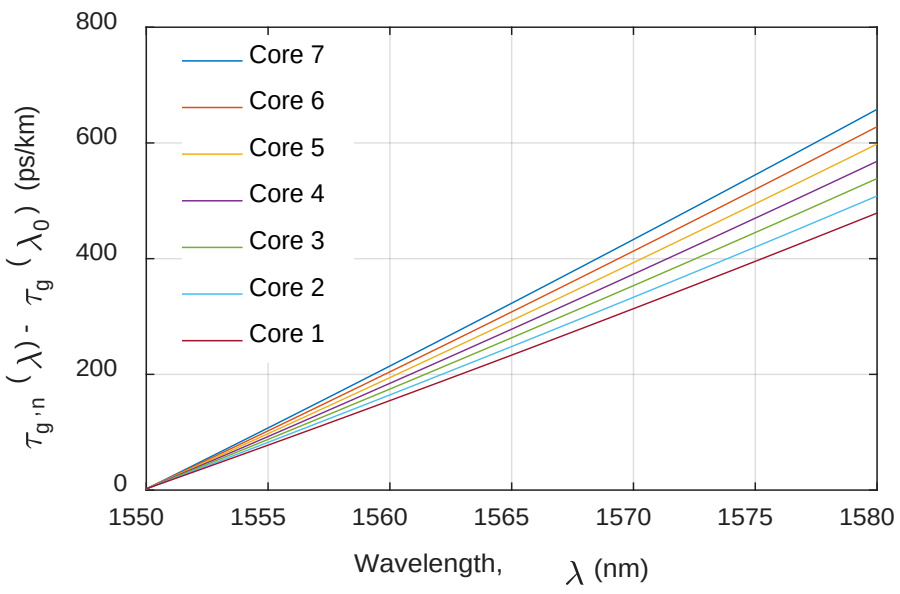

Fig. 2. (a) Cross-section of the designed heterogeneous MCF; (b) Design parameters for each core; (c) Computed spectral group delay for each fiber core.

### *B. Homogeneous Multicore Fibers with inscribed Fiber Bragg Gratings*

The implementation of a sampled TTDL built upon homogeneous MCFs, where all the cores share the same propagation characteristics, requires the in-line incorporation of additional dispersive optical elements. This can be achieved if a multicavity device is created by inscribing individual fiber Bragg gratings (FBGs) at selective positions along the fiber cores. Actually, the inscription of FBGs in SMFs has been widely investigated as dispersive sampled delay lines with 1D operability. If we incorporate the spatial diversity, we can provide 2D operability offering a higher performance versatility. In [6], we fabricated different multicavity TTDL devices using the moving phase mask technique, inscribing either the same grating in planes containing a set of cores or individual gratings in single cores. Figure 3(a) illustrates the experimental setup of a microwave signal filter implemented with a multicavity device based on independent FBG inscription in 3 of the outer cores of a 7-core fiber with a 125-µm cladding diameter and 35-µm core pitch. Each core comprises an array of 3 uniform FBGs centered at different optical wavelengths and located at different longitudinal positions. The gratings in a given core (different wavelengths) were inscribed with incremental distances: 20 mm for core 6, 21 mm for core 5 and 22 mm for core 4. The gratings centered at the same optical wavelength (different cores) were spaced 6 mm for optical wavelength $\lambda_1$, 7 mm for $\lambda_2$ and 8 mm for $\lambda_3$.

Figure 3(b) shows the measured filter frequency responses when we exploit wavelength diversity, (that is, when gathering the samples coming from a given core). We can change the filter free spectral range by selecting one core or another. If we detect instead the signal samples coming from different cores (thus sharing the same optical wavelength), we can operate the delay line using spatial diversity, as shown in Fig. 3(c).

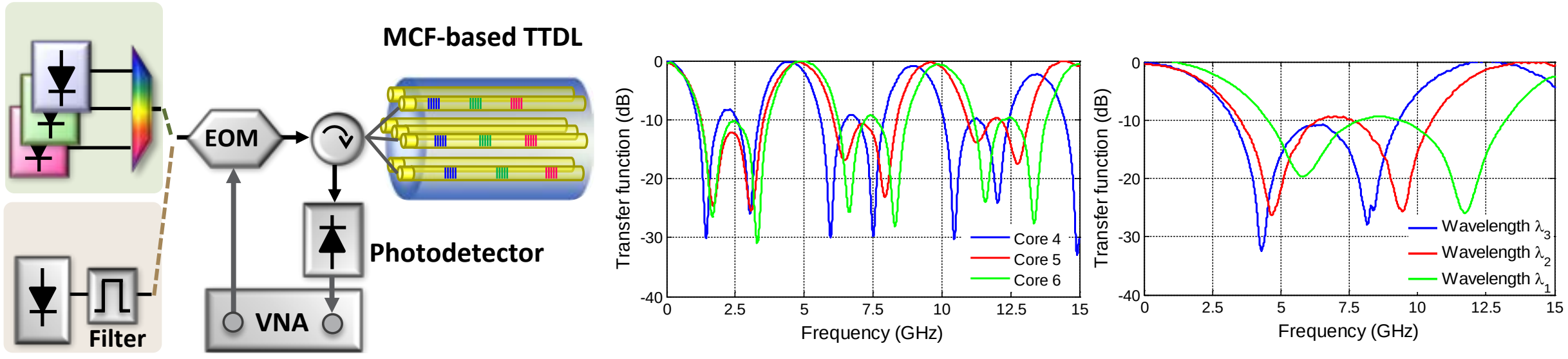

Fig. 3 (a) Experimental setup of a microwave signal filter built upon a homogeneous MCF multicavity device; measured radiofrequency responses operating in (b) wavelength diversity and (c) spatial diversity.

## C. *Few-mode Fibers*

We can also develop distributed signal processing over FMF links if we engineer the fiber so that every mode (or group of modes) experiences, at a given optical wavelength, the adequate group delay to fulfil TTDL operation. We must assure, in addition, a low level of coupling between groups of modes, what calls for the use of refractive step-index profiles and short distances combined with direct detection. In [7], we reported the experimental demonstration of RF signal processing on a step-index 60-m FMF link provided by Prysmian that is characterized by a 15-µm core diameter and a refractive index contrast of 1.1%. At a wavelength of 1550 nm, the differential group delays (relative to $LP_{01}$) are 4.4, 8.9 and 7.9 ps/m, respectively for $LP_{11}$, $LP_{21}$ and $LP_{02}$ modes, and the chromatic dispersion values are 21, 26, 19 and 8 ps/km/nm, respectively for $LP_{01}$, $LP_{11}$, $LP_{21}$ and $LP_{02}$ modes. The inscription of a long period grating (LPG) at a specific position along the fiber converts part of the $LP_{01}$ mode into the $LP_{02}$, permitting sample time delay engineering.

Two 3-sample TTDL configurations with different time delay properties were implemented over the same fiber depending of the modes selected at detection, resulting in a filter with FSR = 3.75 GHz when selecting $LP_{01}$, $LP_{11a}$ and $LP_{21a}$ modes, and a filter with FSR = 7.50 GHz when detecting $LP_{01}$, $LP_{11a}$ and $LP_{02}$ modes. Figure 4(a) illustrates the concept underlying both TTDLs where the LPG is required to control the group delay of $LP_{02}$ mode. Fig. 4(b) shows the normalized RF response of the microwave filter implemented with the TTDL produced by the set of modes $LP_{01}$, $LP_{02}$ and $LP_{11a}$. As we can see, the experimental results show a very good agreement with the theoretical simulation.

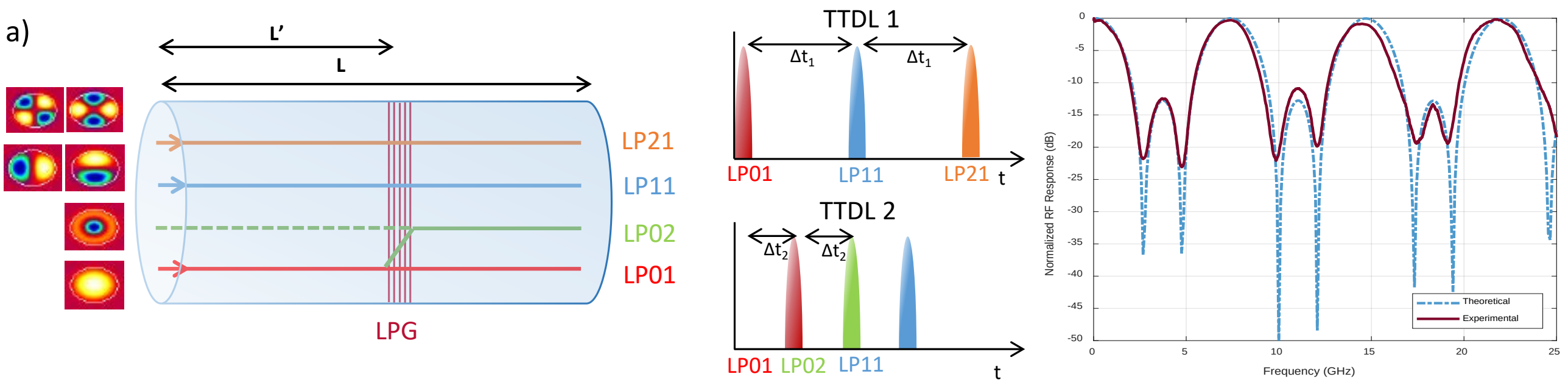

Fig. 4. (a) Sampled TTDL operation over an FMF with an LPG inscribed at the position *L'* to excite the $LP_{02}$ mode; (b) Theoretical and experimental responses of the RF filter implemented with the set of modes: $LP_{01}$, $LP_{02}$ and $LP_{11a}$.

## 3. CONCLUSIONS

Future fiber-wireless access networks will potentially benefit from SDM-based approaches in terms of compactness as compared to a set of parallel SMFs and in terms of operation versatility offered by the simultaneous use of the spatial- and wavelength-diversity domains. We have reported distributed RF signal processing solutions built upon dispersion-engineered heterogeneous MCF links, multicavity devices on commercial homogeneous MCFs and FMF links combined with inscribed gratings for mode conversion and group delay adjustment.

## REFERENCES

[1] D.J. Richardson, J.M. Fini, and L.E. Nelson, "Space-division multiplexing in optical fibres," Nat. Photonics, vol.7, pp. 354-362, 2013.

[2] J. M. Galvé, I. Gasulla, S. Sales, and J. Capmany, "Reconfigurable Radio Access Networks using Multicore Fibers," IEEE J. of Quantum Electronics, vol. 52, pp. 1-7, 2016.

[3] I. Gasulla and J. Capmany, "Microwave photonics applications of multicore fibers," Photonics J., vol. 4, pp. 877-888, 2012.

[4] J. Capmany et al., "Microwave photonic signal processing," IEEE J. Lightw. Technol., vol. 31, no. 4, pp. 571-586, 2013.

[5] S. Garcia, S. and I. Gasulla, "Dispersion-engineered multicore fibers for distributed radiofrequency signal processing," Opt. Express, vol. 24, pp. 20641-20654, 2016.

[6] I. Gasulla, D. Barrera, J, Hervás, and S. Sales, "Spatial Division Multiplexed Microwave Signal processing by selective grating inscription in homogeneous multicore fibers," Sci. Reports, no. 41727, pp. 1-10, 2017.

[7] R. Guillem, S. García, J. Madrigal, D. Barrera, and I. Gasulla, "Few-mode fiber true time delay lines for distributed radiofrequency signal processing," Opt. Express, vol. 26, pp. 25761-25768, 2018.